\documentclass{sigcomm-alternate}
\usepackage{times,url,balance}
\def\sharedaffiliation{%
\end{tabular}
\begin{tabular}{c}}

\begin{document}

\title{
    On Cycles in AS Relationships
}

\numberofauthors{3}
\author{
    \alignauthor Xenofontas Dimitropoulos \\ \affaddr{IBM Research, Z{\"u}rich} \\ \email{xed@zurich.ibm.com}
    \alignauthor M. \'Angeles Serrano \\ \affaddr{IFISC (CSIC-UIB)} \\ \email{marian.serrano@ifisc.uib-csic.es}
    \alignauthor Dmitri Krioukov \\ \affaddr{CAIDA} \\ \email{dima@caida.org}
    \and
    \alignauthor{}
    \sharedaffiliation{\tiny
    This article is an editorial note submitted to CCR. It has NOT been peer reviewed.
    Authors take full responsibility for this article's technical content.
    Comments can be posted through CCR Online.}
}

\maketitle

\category{\\C.2.1}{Network Architecture and Design}{Network topology}

\vspace{.35in}

\textbf{\huge S}everal users of our AS relationship inference
data~\cite{as-rel-data}, released with~\cite{DiKrFo06}, asked us why
it contained AS relationship cycles, e.g., cases where AS $A$ is a
provider of AS $B$, $B$ is a provider of $C$, and $C$ is a provider
of $A$, or other cycle types. Having been answering these questions
in private communications, we have eventually decided to write down
our answers here for future reference.

Formally, the heuristics in~\cite{DiKrFo06} do not produce acyclic
relationships because neither do the techniques that~\cite{DiKrFo06}
is based on~\cite{DiKrHuClRi05,BaEr07,SuAgReKa02}, and because we
did not try to impose any no-cycle constraints that can certainly be
enforced~\cite{CoRa07}. Below we explain why we did not try to do
that, and thus behaved somewhat counter-intuitive.

First, we have to emphasize that AS relationship cycles by no means
imply or induce routing or traffic forwarding loops, since the BGP's
routing loop avoidance mechanism, i.e., AS path checking, does not
depend on any AS relationships. Another problem with BGP causes some
confusion sometimes. BGP can diverge due to policy
conflicts~\cite{GriShe02}, where {\it policy} refers to a ranking of
prefixes that can be used to reach a destination. As AS
relationships determine export policies, they may also influence
prefix rankings, e.g., an AS typically prefers to reach a
destination via its customer, then peer, then provider. However,
this influence by no means leads to BGP divergence, which is
a separate problem with BGP that can only appear in the presence
of non-shortest-path routing and conflicting path selection policies.
AS relationship cycles present or not, the default BGP path
selection mechanism that chooses shortest paths, or many other
possible safe rankings~\cite{GriShe02} result in stable and
loop-free routing.

Second, we have never received a satisfactory answer to our counter
question: Why can the global Internet {\it not} have any AS
relationship cycles? Why does it have to be a DAG (directed acyclic
graph)? AS relationships emerge from business negotiations between
pairs of ASs, while all relevant information is kept secret. In
other words, local negotiations and interactions between AS pairs
determine their relationships. Given the complexity of business
agreements, it is quite unlikely that the local, independent, and
diverse interactions between ASs yield a global, highly organized,
and strictly hierarchical DAG structure. What entity would enforce
and control this structure? Perhaps this entity is ``rational
economy,'' in which there is no money transfer cycles? In stricter
terms, the question is if the Internet market complies with the {\it
efficient market hypothesis}~\cite{Fama70}. We emphasize that there
is no data to answer this question either way. We believe it is
unlikely that the Internet is a perfectly efficient global
market, and refer an interested reader to the criticism of the
efficient market hypothesis in~\cite{GroSti80}.

Another reason why AS relationship cycles may exist in the Internet
is that real AS relationships, especially those between large ISPs,
may be more complex than the course modeling abstraction we adopted
in~\cite{DiKrFo06}. Real AS relationships may depend on a peering
point, prefix, and even time~\cite{DiKrFo06}. For example, ISPs that
dominate one geographical region can provide transit for this region
and receive transit for other regions, thus forming partial transit
relationships with other ISPs. Such geographical interdependencies
clearly lead to relationship cycles: three ASs active primarily in
the US, Europe, and Asia, but also present in the two other regions,
may form a cycle by providing transit to each other for their primary
regions. Interestingly, this kind of information-losing abstractions
of different-type objects or relationships as nodes or links of the same type explain
cycles in other flow networks. For example, in the economic network
of the world trade web~\cite{SerBogVes07}, cycles are present
due to coarse categorizations of products and contractual
relationships. Food webs (nodes are species and directed links show
who eats whom) also have cycles~\cite{AllBodBon05}, quite a
counter-intuitive fact if one considers the biomass flux, but it is
easily explainable by heterogeneities in population composition such
as size, age, etc. Going back to the Internet, the diversity of
real business agreements between ASs is a sufficient condition for
the presence of cycles.

Finally, we did not claim the 100\% accuracy of our results. Our
validation indicated that they were roughly 90\%
accurate~\cite{DiKrFo06}. Incorrect inferences are thus present
in~\cite{as-rel-data}. Suppose that our heuristics mis-infer a
single relationship between a small customer AS $C_1$ and its
provider, large ISP $P$. A single erroneous relationship of this
type can result in many cycles, as all other customers $C_i$,
$i=2,3,\ldots$, of $P$ will then have a path to $C_1$ in the
customer-to-provider direction. Given that $C_1$ can also have other
customer-to-provider paths to $C_i$ via its providers other than
$P$, this type of mis-inference will form many cycles going via $P$,
$C_1$, and $C_i$. Clearly, no-cycle constraints should in theory
suppress this type of errors; we also considered making the
minimization of the number of cycles an objective in our
multiobjective problem formulation, but given all the points above
suggesting that AS relationship cycles may in fact be present in the
real Internet, we decided not to follow this path.

\balance


\scriptsize
\begin{thebibliography}{10}

\bibitem{as-rel-data}
{CAIDA}.
\newblock {AS Relationships Data}.
\newblock Research Project.
\newblock \url{http://www.caida.org/data/active/as-relationships/}.

\bibitem{DiKrFo06}
X.~Dimitropoulos, D.~Krioukov, M.~Fomenkov, B.~Huffaker, Y.~Hyun, kc~claffy,
  and G.~Riley.
\newblock {AS} relationships: Inference and validation.
\newblock {\em Comput Commun Rev}, 37(1), 2007.

\bibitem{DiKrHuClRi05}
X.~Dimitropoulos, D.~Krioukov, B.~Huffaker, kc~claffy, and G.~Riley.
\newblock Inferring {AS} relationships: Dead end or lively beginning?
\newblock In {\em WEA}, 2005.

\bibitem{BaEr07}
G.~D. Battista, T.~Erlebach, A.~Hall, M.~Patrignani, M.~Pizzonia, and
  T.~Schank.
\newblock Computing the types of the relationships between {Autonomous
  Systems}.
\newblock {\em IEEE ACM T Network}, 15(2), 2007.

\bibitem{SuAgReKa02}
L.~Subramanian, S.~Agarwal, J.~Rexford, and R.~H. Katz.
\newblock Characterizing the {Internet} hierarchy from multiple vantage points.
\newblock In {\em INFOCOM}, 2002.

\bibitem{CoRa07}
R.~Cohen and D.~Raz.
\newblock Acyclic type of relationships between {Autonomous Systems}.
\newblock In {\em INFOCOM}, 2007.

\bibitem{GriShe02}
T.~Griffin, F.~B. Shepherd, and G.~Wilfong.
\newblock The stable paths problem and interdomain routing.
\newblock {\em IEEE ACM T Network}, 10(2), 2002.

\bibitem{Fama70}
E.~Fama.
\newblock Efficient capital markets: A review of theory and empirical work.
\newblock {\em J Financ}, 25(2):383--417, 1970.

\bibitem{GroSti80}
S.~J. Grossman and J.~E. Stiglitz.
\newblock On the impossibility of informationally efficient markets.
\newblock {\em Am Econ Rev}, 70(3):393--408, 1980.

\bibitem{SerBogVes07}
M.~\'{A}. Serrano, M.~Bogu{\~{n}}\'{a}, and A.~Vespignani.
\newblock Patterns of dominant flows in the world trade web.
\newblock {\em J Econ Interact Coord}, 2(2):111--124, 2007.

\bibitem{AllBodBon05}
S.~Allesina, A.~Bodini, and C.~Bondavalli.
\newblock Ecological subsystems via graph theory: The role of strongly
  connected components.
\newblock {\em OIKOS}, 110(1):164--176, 2005.

\end{thebibliography}
\end{document}